\documentclass[12pt]{iopart}
\usepackage{graphicx}
\usepackage{epsfig}
\usepackage{amsfonts}
\usepackage{amssymb}

\begin{document} 

\title{Toppling pencils---Macroscopic Randomness from Microscopic Fluctuations}

\author{Thomas Dittrich, Santiago Pe\~na Mart\'\i nez}
\address{Departamento de F\'\i sica, Universidad Nacional de Colombia, Bogot\'a D.C., Colombia}
\ead{tdittrich@unal.edu.co}

\begin{abstract}
We construct a microscopic model to study discrete randomness in bistable systems
coupled to an environment comprising many degrees of freedom. A quartic double well is
bilinearly coupled to a finite number $N$ of harmonic oscillators. Solving the time-reversal
invariant Hamiltonian equations of motion numerically, we show that for $N = 1$,
the system exhibits a transition with increasing coupling strength from integrable to chaotic
motion, following the KAM scenario. Raising $N$ to values of the order of 10 and higher,
the dynamics crosses over to a quasi-relaxation, approaching either one of the stable equilibria
at the two minima of the potential. We corroborate the irreversibility of this relaxation on
other characteristic timescales of the system by recording the time dependences of autocorrelation,
partial entropy, and the frequency of jumps between the wells as functions of $N$ and other parameters.
Preparing the central system in  the unstable equilibrium at the top of the barrier and
the bath in a random initial state drawn from a Gaussian distribution, symmetric under
spatial reflection, we demonstrate that the decision whether to relax into the left or
the right well is determined reproducibly by residual asymmetries in the initial positions
and momenta of the bath oscillators. This result reconciles the randomness and
spontaneous symmetry breaking of the asymptotic state with the conservation
of entropy under canonical transformations and the manifest symmetry of potential
and initial condition of the bistable system.
\end{abstract}

\maketitle

\section{Introduction}\label{sec1}

In the heydays of chaos theory, some 50 years ago, for many a pioneer of this field it came with the
covert or explicit hope that a good part of physical phenomena relegated till then to the realm of
randomness, such as weather and turbulence, could be described comprehensively in terms of
deterministic laws. It rapidly became clear, however, that in many areas, randomness would remain an
indispensable element of the theoretical analysis. Deterministic chaos was combined with noise
\cite{Kap88}, criteria were developed to distinguish chaos from mere chance \cite{KS04},
and with the subject of quantum chaos, the question was addressed how deterministic chaos
could be modified to reconcile it with a theory considered as fundamentally probabilistic.

At the same time, important paradigms of chance in macroscopic phenomena remain that defy an
understanding in terms of deterministic chaos. A fascinating particular instance is randomizing devices
in games of luck, such as tossed coins \cite{DHM07}, dreidels, dice, or roulette wheels \cite{Poi96}.
Their underlying dynamics is not chaotic, it is rather the discretization of the final condition
into two, four, six, or 37 bins that results in a sensitive dependence on the initial condition and
thus reduces a continuous angle coordinate to a practically unpredictable integer.
Notwithstanding, a description in terms of deterministic equations of motion is possible and allows
for example to verify or falsify the presence of biasses in the outcomes.

Much more relevant from a physical point of view are processes that magnify microscopic dynamical disorder in many-body systems to randomness on macroscopic scales. A prototype of this
phenomenon is Brownian motion, where the trajectory of a pollen grain amplifies thermal noise to direct
observability. Traditionally, Brownian motion is represented as a stochastic process \cite{UO30},
using statistical descriptions such as Langevin or Fokker-Planck equations \cite{Ris89},
without any more detailed examination of the underlying microscopic mechanisms.

The present paper intends a synthesis of these two views of randomness, proposing
a model that combines the discreteness of the output with the deterministic dynamics of a
many-body system as random generator on the input side. The macroscopic central component,
representing the tossed coin, is a bistable system, a symmetric double well modelled as a quartic
oscillator. It can be seen as a physical representation of a classical bit, such as an inverted pendulum,
or more graphically even, as a pencil balanced tip down on a flat surface (inset in
Fig.\ \protect\ref{figdoublewell}a).

The microscopic part adheres to the standard modelling of environments as heat baths,
coupling the double well to a set of $N$ harmonic oscillators. It is well known and has been
argued in countless works in statistical mechanics, solid-state physics, and many other fields, that for
$N \to \infty$ and under certain conditions on the frequency dependence of coupling and
spectral density of the oscillators, the bath becomes an irreversible sink of information and energy,
inducing relaxation to a stationary state and dissipation in the central system \cite{Ull66}.
In this point, however, we adopt a more recent development in statistical mechanics,
in that we keep the number of oscillators large, $N \gg 1$, \emph{but finite}
\cite{PC04,SO08,RB08,Has11,JN+13}, so that the dynamics of the total system can be treated
in the framework of the time-reversal invariant Hamiltonian mechanics of closed systems.
It has been demonstrated for classical as well as for quantum systems \cite{GKG10,GBS14},
and is corroborated by the present work, that despite its time-reversal symmetry,
this approach reproduces irreversible behaviour on all relevant timescales. Poincar\'e recurrences,
which prevent true irreversibility in systems with a finite number of freedoms, occur only on timescales
that diverge geometrically with $N$ \cite{MM60}. Our purpose, however, is not substantiating
the approach to thermal equilibrium in these systems. Finite heat baths offer another advantage
we exploit in the present context and which is excluded from the outset in an ensemble treatment:
Fluctuations of the bath now become controllable and reproducible. This allows us to specify
the initial conditions for each oscillator  individually and in this way, to study how these fluctuations
become manifest in the macroscopic randomness of the final state of the central system.

In particular, we would like to demonstrate that the outcome of this game of luck, whether the central
system, initially prepared exactly in a ``Buridan's ass state'', the unstable equilibrium position on top
of the barrier, falls into the left well (``tail'') or the right well (``head''), depends on asymmetries in
the initial condition of the oscillators in the bath. Balance the pencil precisely tip down: If it still falls over,
in which direction will it fall? It is determined by the environment, the particles of the surrounding gas
impinging on the pencil. More generally, the bistable system amplifies and thus measures random
fluctuations in the microscopic degrees of freedom, converting them into random bits. Looking only at
the central bistable system, the random sequence thus generated amounts to a productioncof one bit
of entropy per run of the experiment. Here, another aspect of the Hamiltonian dynamics of closed
systems comes in handy, the conservation of entropy under canonical transformations \cite{Dit19}.
It implies that the entropy in the random sequence cannot be produced by the central system
but must originate somewhere else in the total system. The only possible source is the environment
embodied in the heat bath. From a different point of view, the falling pencil violates
the rotational symmetry with respect to the vertical axis of the total potential, including the interaction
with the environment, and of its own initial condition. The symmetry breaking must therefore occur in
the initial condition of the environment. Finally, with this random bit the system retains a lasting memory,
albeit minimal, of its initial state, a blatant manifestation of its non-Markovian nature.

Deterministic chaos reduces entropy production to the expansion of the initial condition
by the chaotic phase-space flow \cite{Sha81}. To be sure, already for $N = 1$, the quartic double well
coupled to harmonic oscillators is indeed a partially chaotic system. However, this is not decisive
for the randomness exhibited by the bistable system. The pivotal factor is rather
the many-body nature of the bath. In this sense, what we see is Brownian motion discretized
and condensed into random bits.

In fact, this work is inspired and motivated by a similar situation in quantum mechanics. Spin
measurement is a paradigm of irreducible randomness in quantum mechanics, it serves as a source
of binary random numbers less predictable than any classical physical or digital random number
generator, and therefore a valuable resource and a gold standard for applications such as
cryptography \cite{JC+11,BA+17,BK+18}. A quantum two-state system such as a spin-$\frac{1}{2}$
neither has a classical limit nor can it be understood as the quantization of a classical bistable system.
However, the double-well potential is regarded as the closest classical analogue of a qubit,
and the isolated ground-state pair of a quantum double well can be mapped one-to-one
to a qubit \cite{LC+87}. Inhowfar the results of the present work suggest any new insight
concerning the interpretation of quantum randomness is presently under study.

We review the anatomy of the quartic double well in Subsection \ref{sec21}, together with an outline
of the Hamiltonian as well as the dissipative dynamics of this bistable system. Subsection \ref{sec22}
details the construction of the heat bath and sketches some basic facts about the irreversible
relaxation process approached in the limit $N \to \infty$ of the number $N$ of bath modes.
Numerical results confirming and illustrating the chaotic behaviour of the double well coupled
to a single harmonic oscillator, $N = 1$, are presented in Section \ref{sec3}. The central
Section \ref{sec4} is dedicated to our main results providing numerical evidence for
the relaxation of the bistable system into one of its stable equilibrium positions for $N \gg 1$
and the dependence of the final state on the initial condition of the bath. Finally, Section \ref{sec5}
reflects on the implications of our results for our conception of randomness.

\section{The model: bistable system coupled to a finite heat bath}\label{sec2}

A straightforward way of modelling a multistable system is combining a potential with a
corresponding number $n$ of relative minima with a dissipative dynamics, for example Ohmic
friction. In the overdamped regime, where inertia can be neglected against potential forces and
friction, the system will fall from any initial condition into the closest well. We here follow this
simple scheme to model a bistable system, i.e., for $n = 2$, to be construed in Subsection
\ref{sec21}, before coupling it to a finite heat bath in Subsection \ref{sec22}.

\subsection{Quartic double well}\label{sec21}

\begin{figure}[h!]
\begin{center}
\includegraphics[width=11 cm]{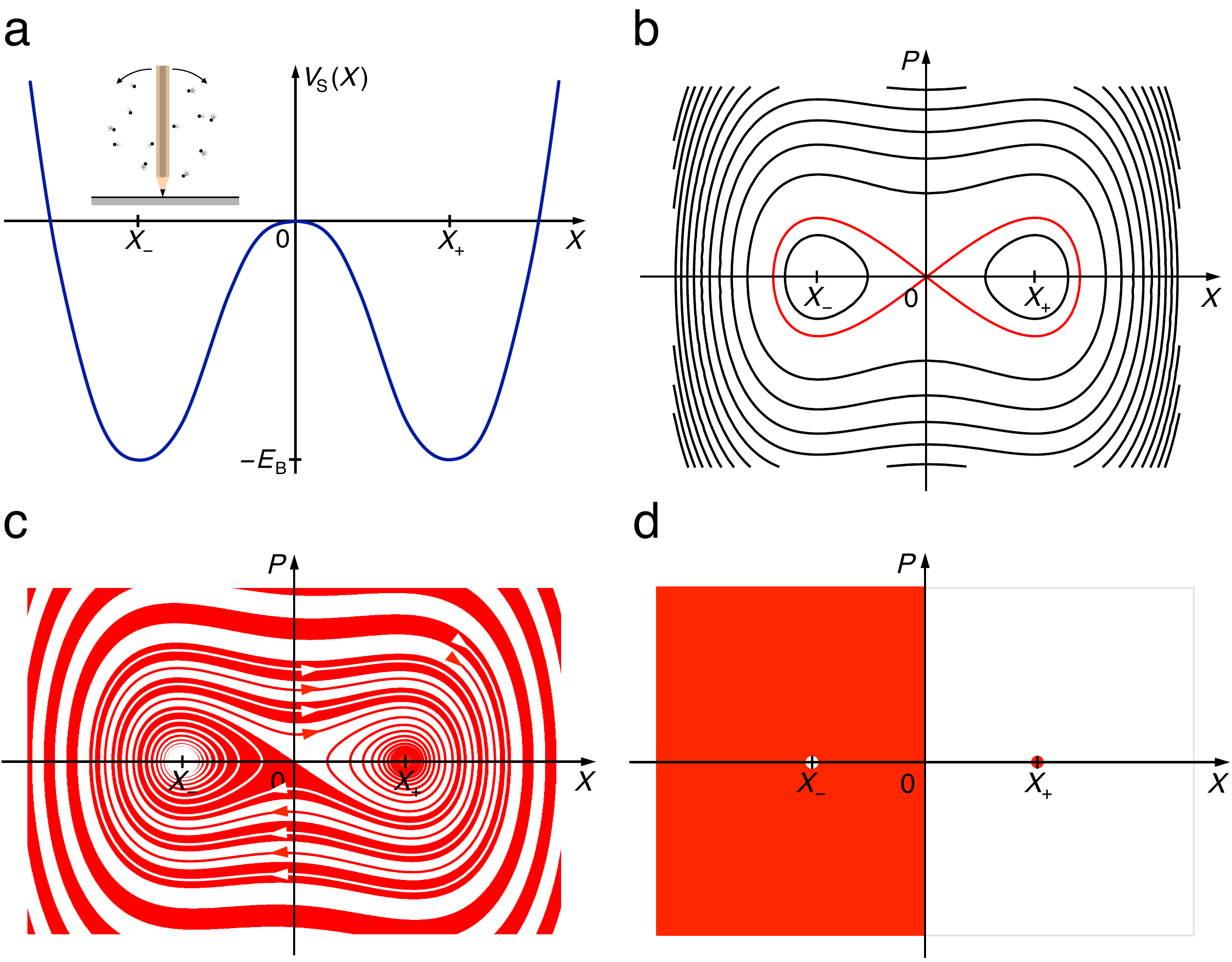}
\caption{
Hamiltonian motion (panels a,b), Eqs.\ (\protect\ref{dwpotential},\protect\ref{dwhamiltonian}),
and dissipative dynamics (c,d), Eq.\ (\protect\ref{dwnewton}), for a quartic double well. 
The potential (a), modelling a pencil balanced tip down on a flat surface (inset),
shows two quadratic minima at $X_\pm$, related by parity $X \to -X$,
separated by a parabolic barrier with its top at $X_0 = 0$. Trajectories of the
Hamiltonian dynamics (b) comprise approximately harmonic oscillations within each well
for negative and strongly anharmonic oscillations for positive energies, separated by
a separatrix (red) at $E = 0$. Basins of attraction for weak friction (c) spiral around
one another into either well. In the limit of strong friction (d), they coincide with
the half spaces $X < 0$ and $X > 0$.}
\label{figdoublewell}
\end{center}
\end{figure}

\noindent
Modelling a bistable system, any potential with two symmetry-related minima will do,
but for the sake of mathematical transparency and ease of calculation, we prefer
the standard potential  of a quartic oscillator with a parabolic barrier (Fig.\ \protect\ref{figdoublewell}a),
\begin{equation} \label{dwpotential}
V_{\rm{S}}(X) = - \frac{a}{2} X^2 + \frac{b}{4} X^4 ,\quad a,b \in \mathbb{R}^+ .
\end{equation}
It has quadratic minima at $X_\pm = \pm \sqrt{a/b}$ and a quadratic maximum at
$X_0 = 0$ of relative height $E_{\rm B} = a^2/4b$. The only relevant parameter for the shape
of this potential is the relative sign of $a$ and $b$, even a variation of the ratio $a/b$ can be
compensated by a corresponding rescaling of position or energy or both.

Frictionless motion in this potential (Fig.\ \protect\ref{figdoublewell}b) is described by the Hamiltonian
\begin{equation} \label{dwhamiltonian}
H(P,X) = \frac{P^2}{2M} + V(X).
\end{equation}
Close to the minima at $X_\pm$, it consists of harmonic oscillations with the frequency
$\Omega = \sqrt{2a/M}$. They become increasingly anharmonic as the energy rises towards the top
of the barrier. At $E = 0$, on the level of the barrier top, the trajectory assumes a figure-8 shape,
the separatrix. Close to the top, the dynamics is governed by an unstable manifold,
along which distances in phase space increasing exponentially with the Lyapunov exponent
$\Lambda = \sqrt{a/M}$, and a stable manifold contracting phase space correspondingly.
At higher energies $E > 0$, oscillations are strongly anharmonic
and circle both wells, passing over the barrier back and forth.

Remaining on the macroscopic level of description, dissipation is included as a damping term
in Newton's equations of motion, as derived otherwise from the Hamiltonian (\ref{dwhamiltonian}),

\begin{equation} \label{dwnewton}
M \ddot X = a X - b X^3 - 2\Gamma \dot X,
\end{equation}

\noindent
assuming Ohmic friction with damping coefficient $2\Gamma$. Solutions of Eq.\ (\ref{dwnewton})
now contract phase space exponentially with a rate $2\Gamma$ towards a pair of point attractors,
one at the bottom of each minimum. In the underdamped regime, for $\Gamma < \Omega$, from an
initial energy $E(0) < 0$, below the top of the barrier, trajectories spiral from either side of
the barrier into the adjacent well, the basins of attraction forming a Yin-and-Yang figure
(Fig.\ \protect\ref{figdoublewell}c). At higher energies, they wind out around one another
and around the two wells. In the overdamped regime, these spirals get steeper, the basins of attraction approaching the half spaces $X \in {\mathbb{R}}^-$ for $X_-$ and $X \in {\mathbb{R}}^+$ for $X_+$,
resp. (Fig.\ \protect\ref{figdoublewell}d)

In all regimes, independently of the degree of friction, the equations of motion are solved by
$(P,X) = \rm{const} = (0,0)$ if the system is prepared at rest on top of the barrier.
This is an isolated point in phase space. With nonzero
friction, any infinitesimal deviation from this unstable equilibrium will send the system
into one of the wells. Even in the Hamiltonian case, trajectories started on the separatrix
do not pass through this point but need an infinite time to approach it.

\subsection{Finite heat bath}\label{sec22}

Coupling the bistable system to an environment with a large number of degrees of freedom
requires including three terms in the Hamiltonian,

\begin{equation} \label{sseehamiltonian}
H(\mathbf{R},\mathbf{r}) =
H_{\rm{S}}(\mathbf{R}) + H_{\rm{SE}}(\mathbf{R},\mathbf{r}) + H_{\rm{E}}(\mathbf{r}),
\end{equation}

\noindent
$\mathbf{R} = (P,X)$ denoting the phase-space coordinates of the central system and
$\mathbf{r} = (p_1,p_2,\ldots,p_N,$ $x_1,x_2,\ldots,x_N)$ those of the environment comprising $N$
degrees of freedom. The self-energy of the central system is given by the quartic double well,
Hamiltonian (\ref{dwhamiltonian}). For the environment we choose a set of $N$ harmonic oscillators,

\begin{equation} \label{ehamiltonian}
H_{\rm{E}}(\mathbf{r}) = \sum_{n=1}^N \left( \frac{p_n^2}{2m} + \frac{m\omega_n^2}{2} x_n^2 \right).
\end{equation}

\noindent
The frequencies $\omega_n$, $n = 1,\ldots,N$, will be specified further below. Every oscillator
should exert a force, constant in space, on the central system. This suggests to model
their interaction as a linear position-position coupling,

\begin{equation} \label{sehamiltonian}
H_{\rm{SE}}(\mathbf{R},\mathbf{r}) = H_{\rm{SE}}(X,\mathbf{x}) = -X \sum_{n=1}^N g_n x_n,
\end{equation}

\noindent
with coupling constants $g_n$, $n = 1, \ldots ,N$. It does not break
the invariance of the total system under parity (spatial reflection) P: $(\mathbf{r},\mathbf{R}) \to
(-\mathbf{r},-\mathbf{R})$. However, it drives the two minima apart, from $X_{\pm} = \pm \sqrt{a/b}$
to $X_{\pm} = \pm \sqrt{\frac{1}{b}(a+\sum_{n=1}^N g_n^2/2m \omega_n^2)}$,
see Fig.\ \protect\ref{figdwhopot}b, an effect not intended with the coupling to an environment.
It can be compensated for by including a counter term $\sim X^2$ in the potential to complete
the squares with respect to the dependence on the oscillator coordinates,
see Fig.\ \protect\ref{figdwhopot}c,

\begin{eqnarray} \label{sepotential}
V_{\rm{SE}}(X,\mathbf{x}) &= V_{\rm{S}}(X) + \sum_{n=1}^N \frac{m\omega_n^2}{2} x_n^2 -
X \sum_{n=1}^N g_n x_n + X^2 \sum_{n=1}^N \frac{g_n^2}{2m\omega_n^2} \\
&= V_{\rm{S}}(X) + \sum_{n=1}^N \frac{m\omega_n^2}{2} \left(x_n - \frac{g_n}{m\omega_n^2}X\right)^2.
\end{eqnarray}

\noindent
Without pretending any kind of rigorous quantization, we consider the Hamilton operator

\begin{eqnarray} \label{spinbosonhamiltonian}
\hat H &= \hat H_{\rm{S}} + \hat H_{\rm{SE}} + \hat H_{\rm{E}}, \quad
\hat H_{\rm{S}}  = \frac{1}{2} \hbar \Omega \hat \sigma_x, \quad
\hat H_{\rm{E}} = \sum_{n=1}^N \hbar \omega_n \left(\hat a_n^\dagger \hat a_n +
\frac{1}{2} \right), \nonumber \\ 
\hat H_{\rm{SE}} &= \sum_{n=1}^N \hat \sigma_z \bigl(g_n\hat a_n^\dagger + g_n^*\hat a_n \bigr),
\end{eqnarray}

\noindent
with Pauli spin matrices, $\hat \sigma_x$ and $\hat \sigma_z$, and boson creation and annihilation
operators, $\hat a_n^\dagger$ and $\hat a_n$, resp., known as \emph{spin-boson model}
\cite{Lou73,LC+87}, as a close quantum analogue of the classical Hamiltonian
(\ref{sseehamiltonian}) to (\ref{sepotential}).

In the limit $N \to \infty$ of a quasicontinuous spectrum of the bath oscillators and under certain
conditions on the spectral density and the frequency dependence of the coupling, the dynamics
of the central system exhibits irreversible relaxation into a stable state, superposed with
stochastic fluctuations. To be more precise, a decisive quantity is the coupling strength function
\cite{Ull66}, defined by

\begin{equation} \label{couplingstrength}
\gamma(\omega) {\rm{d}} \omega :=
\sum_{\scriptstyle n \atop \scriptstyle \omega \le \omega_n \le \omega + {\rm{d}} \omega} g_n^2.
\end{equation}

\noindent
Close to either one of the quadratic minima of the double well, if the total coupling is not too strong,

\begin{equation} \label{couplingbound}
G^2 := \int_0^\infty {\rm{d}}\omega \frac{\gamma(\omega)}{\omega^2} \le \Omega^2,
\end{equation}

\noindent
the central system behaves as a harmonic oscillator subject to Ohmic friction and
a fluctuating force. The equations of motion for $X(t)$ in general take the form of
integro-differential equations with integral kernels that are nonlocal in time \cite{Ull66}.
However, if $\omega^{-2} \gamma(\omega)$ is approximately constant within a frequency range
$\Delta \omega$ containing $\Omega$, the autocorrelation time of the bath responsible
for the memory, reduces as $\tau_{\rm{E}} \sim \Delta \omega^{-1}$, and for
$\Delta \omega \gg \Omega$, the response of the bath decays instantaneously.
For much larger frequencies, $\gamma(\omega)$ must be cut off, say exponentially
$\gamma(\omega) \sim \exp (-\omega/\omega_{\rm{co}})$, in order to satisfy
Eq.\ (\ref{couplingbound}).

Under these assumptions, the dynamics of the central system is described by a Langevin equation
\cite{UO30,Ull66,Ris89},

\begin{equation} \label{wellangevin}
\ddot X(t) + 2 \Gamma \dot X(t) + \Omega'^2 X(t) = \frac{1}{M} \tilde F(t),
\end{equation}

\noindent
with a friction coefficient $\Gamma = \pi \gamma(\Omega) /4 \Omega^2$ and a modified frequency
$\Omega' = \sqrt{\Omega^2 - \Gamma^2}$. The fluctuating force $\tilde F(t)$ with zero mean,
$\langle \tilde F(t) \rangle = 0$, is delta-correlated,

\begin{equation} \label{bathcorr}
\langle \tilde F(t) \tilde F(t+s) \sim \delta(s).
\end{equation}

In the vicinity of the barrier top, similar considerations apply, but with the natural frequency
$\Omega = \sqrt{2a/M}$ of oscillations in the wells replaced by the Lyapunov exponent
associated to the parabolic barrier, taken as an imaginary frequency, $\Omega \to
{\rm{i}} \Lambda = {\rm{i}} \sqrt{a/M}$. The Langevin equation analogous to
 Eq.\ (\ref{wellangevin}), valid in this neighbourhood, therefore reads

\begin{equation} \label{toplangevin}
\ddot X(t) + 2 \Gamma \dot X(t) - \Lambda'^2 X(t) = \frac{1}{M} \tilde F(t),
\end{equation}

\noindent
and is solved by trajectories expanding or contracting with the rates $-\Gamma \pm \Lambda'$,
$\Lambda' = \sqrt{\Lambda^2 + \Gamma^2}$, along the unstable and stable manifolds
in phase space, resp., emanating from the top of the barrier. In terms of the interplay of
Eq.\ (\ref{toplangevin}), valid near the maximum, and Eq.\ (\ref{wellangevin}), valid near
the minima, it is the initial amplification of the fluctuating force $\tilde F(t)$ along the unstable
manifold, frozen in and reduced to either one of two asymptotic states for $t \gg \Gamma^{-1}$,
that interests us here. Details of the way the bistable system coupled to a finite bath approaches
these states will be discussed in Subsections \ref{sec42} and \ref{sec43}.

\section{Double well coupled to a single harmonic oscillator or a few of them: chaotic dynamics}\label{sec3}

An important aspect of our model, to be contrasted with the regime $N \gg 1$ of a bath
comprising a large number of degrees of freedom, is the case $N = 1$ of a quartic double well
coupled to a single harmonic oscillator. With its two degrees of freedom, it is still far from even any
symptoms of relaxation. However, involving strong anharmonicity in one of its freedoms, is meets all
conditions to become chaotic for non-zero coupling. In this section, we present numerical evidence
that this is indeed the case.

\begin{figure}[h!]
\begin{center}
\includegraphics[width=11 cm]{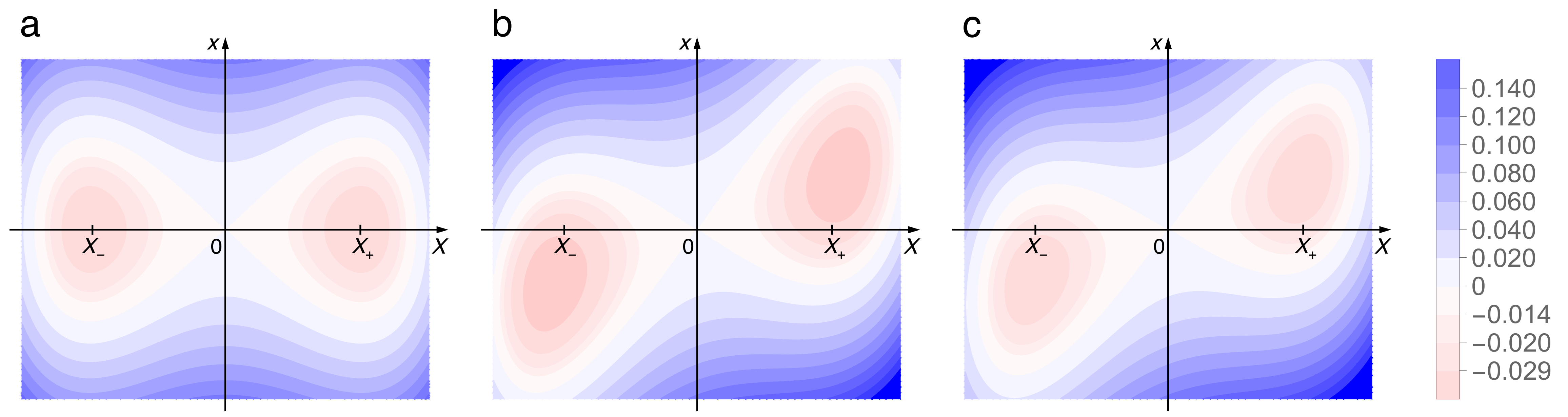}
\caption{
Contour plots of the potential (\protect\ref{dwonebosonpotential}) with parameters
$a = b = 0.15$, $m = 1$, $\omega = 0.4$, without coupling (a), with coupling $g = 0.06$,
but not including the counter term $g^2 X^2 / (2m\omega^2)$ (b), and with this term (c).
Colour code ranges from red (negative) through white (zero) through blue (positive).
}
\label{figdwhopot}
\end{center}
\end{figure}

The total Hamiltonian for $N = 1$ reads,

\begin{equation} \label{dwonebosonhamiltonian}
H(\mathbf{R},\mathbf{r}) = \frac{P^2}{2M} +  \frac{p^2}{2m} + V_{\rm{SE}}(X,x),
\end{equation}

\noindent
where (Fig.\ \protect\ref{figdwhopot})

\begin{eqnarray} \label{dwonebosonpotential}
V_{\rm{SE}}(X,x) &= - \frac{a}{2} X^2 + \frac{b}{4} X^4 + \frac{m\omega^2}{2} x^2 -
g x X + X^2 \frac{g^2}{2m\omega^2} \nonumber \\
  &=  - \frac{a}{2} X^2 + \frac{b}{4} X^4 + \frac{m\omega^2}{2}\left(x - \frac{g}{m\omega^2}X\right)^2, 
\end{eqnarray}

\noindent
It is invariant under parity, $(P,X,p,x) \to (-P,-X,-p,-x)$,
but to our best knowledge lacks any other symmetry or constant of motion.
At the same time, both subsystems are separately integrable for $g = 0$, so that we expect
to see a generic Kolmogorov-Arnol'd-Moser (KAM) \cite{LL83} scenario in
the transition from purely regular to strongly chaotic motion with increased.

In all regimes, independently of the degree of friction, a solution of the equations of motion is
$(P,X) = \rm{const} = (0,0)$ if the system is prepared in this state on top of the barrier.
This is an isolated point in phase space. With nonzero
friction, any infinitesimal deviation from this unstable equilibrium will send the system
into one of the wells. In the Hamiltonian case, even trajectories started on the separatrix
never reach this point but need an infinite time to approach it.

Numerical solutions of Hamilton's equations of motion with the Hamiltonian
(\ref{dwonebosonhamiltonian},\protect\ref{dwonebosonpotential}) have been obtained with
a symplectic integration routine based on a first-order Verlet Leapfrog algorithm
\cite{GNS94,RN17,SC18}. Figure \protect\ref{figdwhopot}a shows contour lines of
the potential $V_{\rm{SE}}(X,x)$, without coupling (Fig.\ \protect\ref{figdwhopot}a)
and for $g = 0.06$, without counterterm (Fig.\ \protect\ref{figdwhopot}b) and with it
(Fig.\ \protect\ref{figdwhopot}c). The parity or, equivalently in two dimensions,
$C_2$-symmetry is evident.

\begin{figure}[h!]
\begin{center}
\includegraphics[width=11 cm]{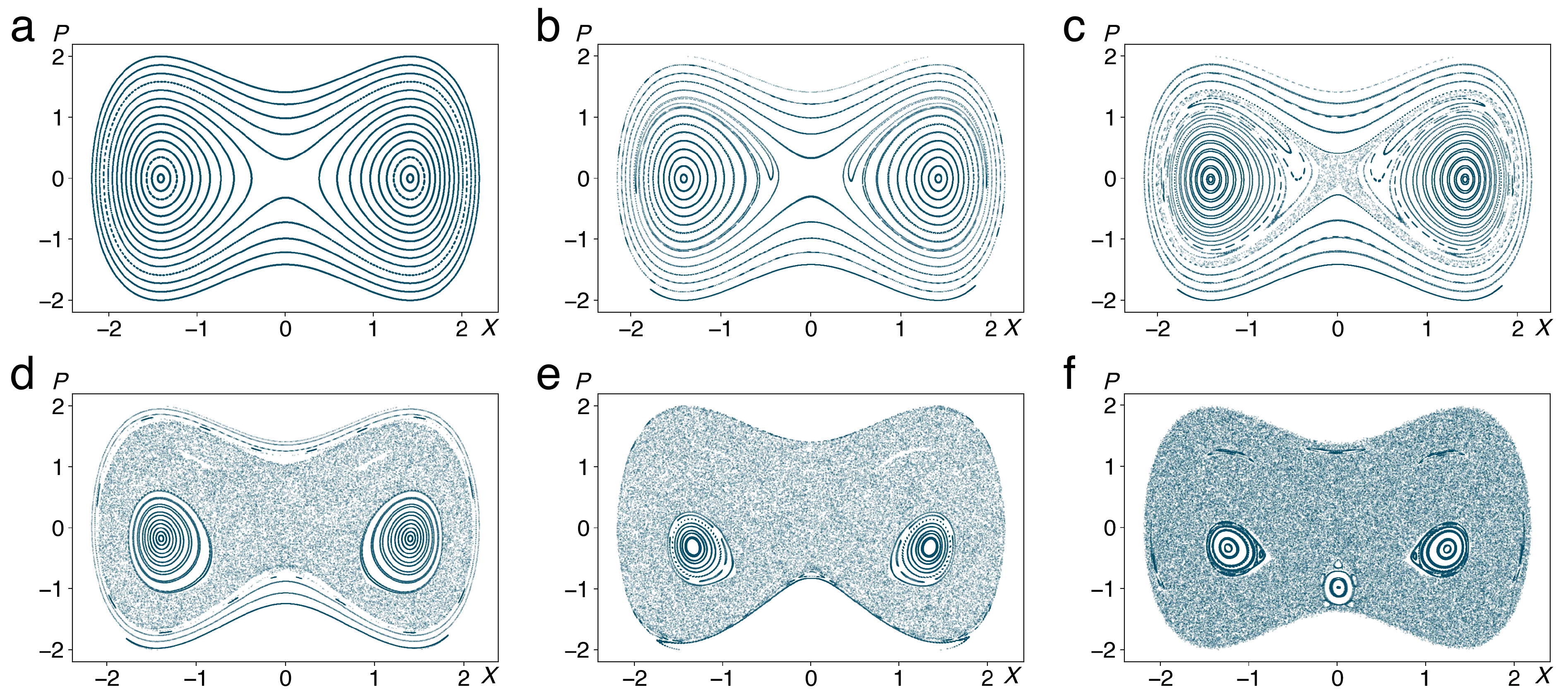}
\caption{
Poincar\'e surfaces of section for the motion generated by the Hamiltonian (\protect\ref{dwonebosonhamiltonian},\protect\ref{dwonebosonpotential}), showing intercepts $(P,X)$
with the hyperplane $x = 0$ under the condition $p > 0$, for $g = 0$ (a), $g = 0.001$ (b), 
$g = 0.005$ (c), $g = 0.05$ (d),  $g = 0.1$ (e),  $g = 0.2$ (f). Other parameters are $a = 2$, $b = 1$,
$M = 1$, $m = 0.1$, $\omega = 1.5$.
}
\label{figpoincaress}
\end{center}
\end{figure}   

In order to visualize trajectories of the system, we use Poincar\'e surfaces of section \cite{LL83,Ott02}
to reduce the three dimensions of the energy shell, the invariant manifold containing the trajectories
within the four-dimensional phase space, further to two. Coordinates $\bigl(P(t),X(t)\bigr)$
are registered whenever trajectories intersect the plane $x = 0$ with $p > 0$ at times $t_n$
(not necessarily equidistant), generating discrete point sequences $(P_n,X_n)$. Surfaces of section for
different values of the coupling constant $g$ are presented in (Fig.\ \protect\ref{figpoincaress}).
For $g = 0$ (Fig.\ \protect\ref{figpoincaress}a), point patterns follow one-dimensional curves
that coincide with the contours of the potential, Fig.\ \protect\ref{figdwhopot}a. Increasing the
coupling to $g = 0.001$ (b) and further to $g = 0.005$ (c), we see irregular motion invading phase space
in the vicinity of the separatrix, in the form of chains of regular islands surrounded by chaotic
regions. The total phase-space area occupied by chaotic trajectories expands further into
the wells and the region above the barrier, resembling a Venetian half mask, as $g$
increases to $0.05$ (d) and $0.1$ (e), till the remaining regular regions reduce to small
islands around the bottoms of the two wells, at $g = 0.2$ (f). The small island visible at
the ``mouth of the mask'', near $(P,X) = (-1,0)$, pertains to a stable periodic trajectory
that follows roughly one of the contours of the potential, see Fig.\ \protect\ref{figdwhopot},
at $V(X,x) > 0$, and represents a surprising effect of the strongly nonlinear dynamics.
With this behaviour, the system follows the well-known KAM scenario.

\begin{figure}[h!]
\begin{center}
\includegraphics[width=11 cm]{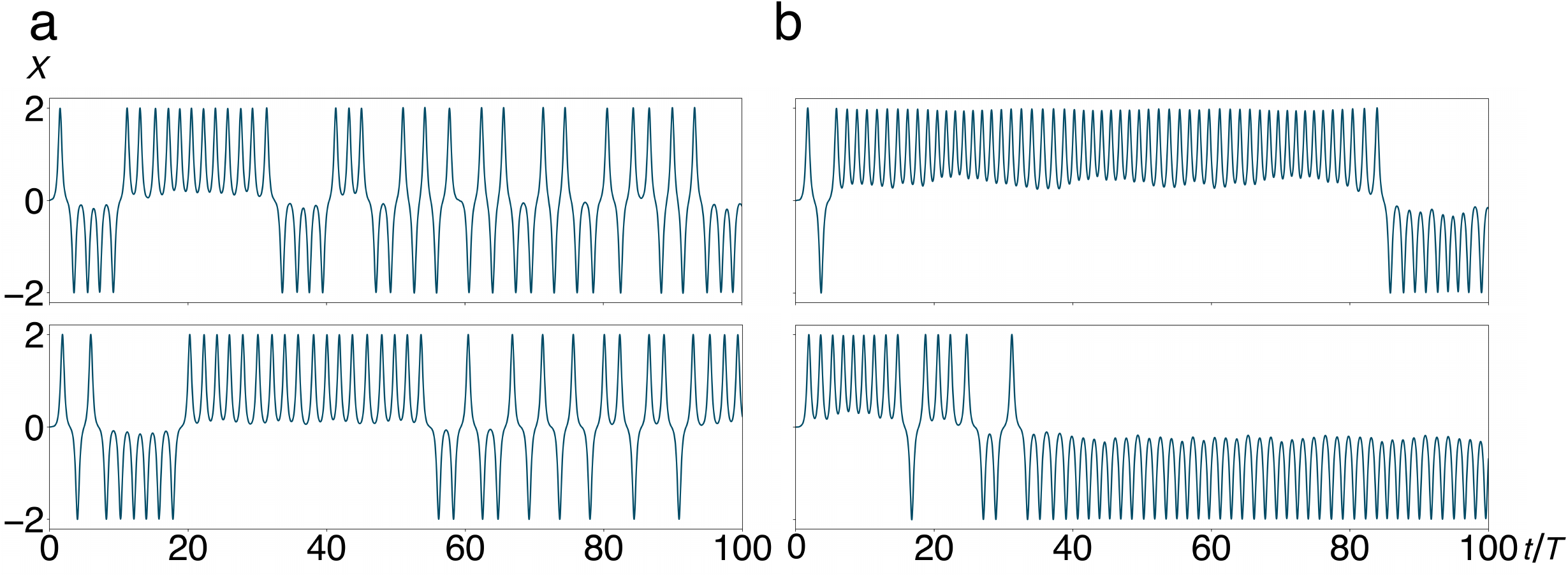}
\caption{
Trajectories $X(t)$ of the quartic double well coupled to a small number $N$ of harmonic oscillators,
Eqs.\ (\ref{dwpotential},\ref{dwhamiltonian},\ref{sseehamiltonian}-\ref{sepotential}), launched from
an initial state at rest on top of the barrier, $\bigl(P(0),X(0)\bigr) = (0,0)$,
show sporadic transitions (``jumps'') between the wells, for $N = 1$ (a) and $N = 5$ (b).
Time is measured in units of the period $T = 2\pi / \Omega$ of unperturbed oscillations
around the stable equilibria of the double well. Other parameter values are
$M = 1$, $m = 1$, $g = 0.1$ and $\omega = 1.5$. For $N = 5$, frequencies and initial conditions
of the bath oscillators have been drawn from random distributions, see Subsection \protect\ref{sec41}.
}
\label{figjumps}
\end{center}
\end{figure}   

With more oscillator modes added to the environment, the total system does of course
not return to integrability. Yet, in a different sense, the dynamical disorder does reduce:
The frequency of jumps between the two wells (analogous to spin flips in the quantum
mechanical context) diminishes with increasing $N$. In Fig.\ \protect\ref{figjumps}, we
depict sample trajectories, representing the position $X(t)$ of the central system alone, for
$N = 1$ (a) and 5 (b). We observe a tendency that the frequency of jumps decreases and
the duration of localized episodes, i.e., periods where the system remains in one of the two wells, increases. It can be roughly explained by the fact that an increasing fraction of
the total energy of the system is absorbed by the bath oscillators, so that most of the time,
the central system does not have enough energy to surmount the barrier.

\section{Double well coupled to a large bath: relaxation and localization at random}\label{sec4}

Increasing $N$ further, we approach the regime where it is more appropriate to treat
the bath modes statistically. Keeping their number finite, though, we are invariably
dealing with discrete distributions of frequencies, coupling strengths, etc., and thus
have to be more specific than working with ensembles defined completely by smooth
probability densities. In the incipient field of finite baths, a few strategies, mostly
for the context of quantum systems, have been developed to cope with this situation
\cite{HW+19}, some of which we adopt in the present work.

\subsection{General setup of numerical simulations}\label{sec41}

Basic data on the dynamics of a quartic double well coupled to an environment comprising
$N$ harmonic oscillators, defined by Eqs.\
(\ref{dwpotential},\ref{dwhamiltonian},\ref{sseehamiltonian}-\ref{sepotential}),
are gained by solving Hamilton's equations of motion,

\begin{eqnarray} \label{sehamiltonseqm}
\dot P &= a X - b X^3 + \sum_{n=1}^N g_n x_n, \quad \dot X = \frac{P}{M}, \\
\dot p_n &= -m \omega_n^2 x + g_n X, \quad \dot x_n =  \frac{p_n}{m}, \quad n = 1,\ldots,N,
\end{eqnarray}

\noindent
using a Calvo-Sanz-Serna $4^{\rm{th}}$ order symplectic integrator \cite{GNS94,RN17,SC18}.
A guideline for the definition of frequencies $\omega_n$ and couplings $g_n$ are the conditions,
mentioned in Section \ref{sec22}, for Ohmic friction and a delta-correlated fluctuating force.
They suggest to choose the coupling strength function (\ref{couplingstrength}) as

\begin{equation} \label{discgammaco}
\gamma(\omega) = \sum_{n=1}^N g_n^2 \delta(\omega - \omega_n)
\sim \omega^2 \exp\left(-\frac{\omega}{\omega_{\rm{co}}}\right).
\end{equation}

\noindent
including an exponential cutoff at $\omega_{\rm{co}}$. In the context of quantum decoherence
and dissipation, the relevant quantity considered instead of $\gamma(\omega)$ is
the so-called \emph{spectral function} \cite{Ull66,LC+87,HW+19},

\begin{equation} \label{spectralfunction}
J(\omega) := \frac{\pi}{2} \sum_{n} \frac{g_n^2}{m\omega_n} \, \delta(\omega - \omega_n).
\end{equation}

\noindent
Writing the spectral function as a product

\begin{equation} \label{couplingspect}
J(\omega) = f\bigl(g(\omega)\bigr) \rho_\omega(\omega),
\end{equation}

\noindent
makes it explicit that it combines the effects of the frequency dependence
$f\bigl(g(\omega)\bigr)$ of the coupling with the pure density of states $\rho_\omega(\omega)$.
For this product, the frequency dependence equivalent to Eq.\ (\ref{discgammaco}) is

\begin{equation} \label{ohmicspectralfunction}
J(\omega) \sim \omega \exp\left(-\frac{\omega}{\omega_{\rm{co}}}\right).
\end{equation}

\noindent
We assemble sets of $N$ harmonic oscillators that satisfy Eq.\ (\ref{discgammaco})
by adapting only the density of states to this condition while keeping the couplings constant
within the bath for given $N$. In order to maintain the total interaction energy independent
of the number of degrees of freedom in the bath, we scale the couplings globally with $N$ as

\begin{equation} \label{couplings}
g_n = {\rm{const}} = \frac{g}{\sqrt{N}}, \quad n = 1,\ldots,N,
\end{equation}

\noindent
with a global system-bath coupling $g$. At the same time, we define a sequence of discrete
frequencies $\omega_n$, not equidistant but with variable frequency steps adjusted such that
the resulting spectral density satisfies Eq.\ (\ref{discgammaco}), see Appendix \ref{specdisdens}.
For an Ohmic coupling strength function $\gamma(\omega) \sim \omega^2$ for
$\omega \ll \omega_{\rm{co}}$, this implies to discretize the frequencies as
(Fig.\ \protect\ref{figspecdisdens}a)

\begin{equation} \label{ohmicdisc}
\omega_n = \Delta \omega \sqrt{2n}, \quad \omega \ll \omega_{\rm{co}}.
\end{equation}

\noindent
The exponential cutoff on a frequency scale $\omega_{\rm{co}}$ included in 
Eq.\ (\ref{discgammaco}) is achieved by a discretization (Fig.\ \protect\ref{figspecdisdens}c)

\begin{equation} \label{cutoffdisc}
\omega_n = - \omega_{\rm{co}} \ln(N_{\rm{co}} - n), \quad \omega \gg \omega_{\rm{co}},
\end{equation}

\noindent
choosing the parameter $N_{\rm{co}}$ according to the desired maximum frequency
$\omega_N = - \omega_{\rm{co}} \ln(N_{\rm{co}} - N)$.



The initial state of the environment playing a central r\^ole for our reasoning, we have
to treat the initial conditions in phase space of the $N$ harmonic oscillators with
particular care. We consider both position $x_n$ and momentum $p_n$ of each oscillator as
Gaussian random variables, defined by probability density functions

\begin{eqnarray} \label{pxgaussian}
\rho_p(p_n) &= \frac{1}{\sqrt{\pi E_p}} \exp\left(- \frac{1}{E_p} \frac{p_n^2}{2m}\right), \\ 
\rho_x(x_n) &= \frac{1}{\sqrt{\pi E_x}} \exp\left(- \frac{1}{E_x} \frac{m\omega_n^2}{2} x_n^2\right).
\end{eqnarray}

\noindent
Independent variances $E_p$ and $E_x$ for momentum and position preserves us
the freedom to vary the aspect ratio of the resulting Gaussian clouds in phase space,
but in most cases, we fix the widths such that $E_p = E_x =: E_{\rm{ho}}$.
The total initial energy in the bath will be kept constant,

\begin{equation} \label{etotal}
E_{\rm{bath}} = \sum_{n=1}^N E_n
\end{equation}

\noindent
mostly at a fraction of the barrier height, so that the individual
initial energies scale on average as $E_n \sim E_{\rm{bath}}/N$. In order that they comply
with Eq.\ (\ref{etotal}) exactly, the $E_n$ are adapted to the required value of $E_{\rm{bath}}$
by scaling all initial positions and momenta accordingly by the same factor
$\sqrt{E_{\rm{bath}} / \sum_{n=1}^N E_n}$.

It is tempting to interpret the densities (\ref{pxgaussian}) as Boltzmann distributions,
defining a temperature through $E_{\rm{ho}} = k_{\rm{B}}T$. However, in view of
the wider scope of this work towards randomness of any origin, we avoid a narrow
interpretation in thermodynamical terms and consider Gaussian distributions
as in Eq.\ (\ref{pxgaussian}) as a practical, rather than compelling, choice.

\subsection{Relaxation to stationary states}\label{sec42}

In order to demonstrate that the double well coupled to a finite environment,
for sufficiently large values of $N$, does approach states that are nearly stationary over long
timescales, we refer to different diagnostics of irreversible behaviour, some of more local,
some of more global character. An appropriate indicator of the loss of memory is
the autocorrelation as a function of the time shift \cite{Ull66,Ris89}. For the position of
the central system, it is defined as

\begin{equation} \label{xxautocorr}
\chi_{XX}\left(t,t + s\right) = 
\frac{\left\langle \bigl(X(t) - \langle X\rangle\bigr)
\bigl(X(t + s)  - \langle X\rangle\bigr)\right\rangle} {\langle X^2 \rangle}.
\end{equation}

\noindent
The angle brackets denote averaging over an ensemble of baths as indicated above.
Moreover, for each configuration of the bath, after transients have decayed, the process
can be considered stationary and we can also average over time $t$ in each time series, 
keeping the time shift $s$ constant.

\begin{figure}[h!]
\begin{center}
\includegraphics[width=11 cm]{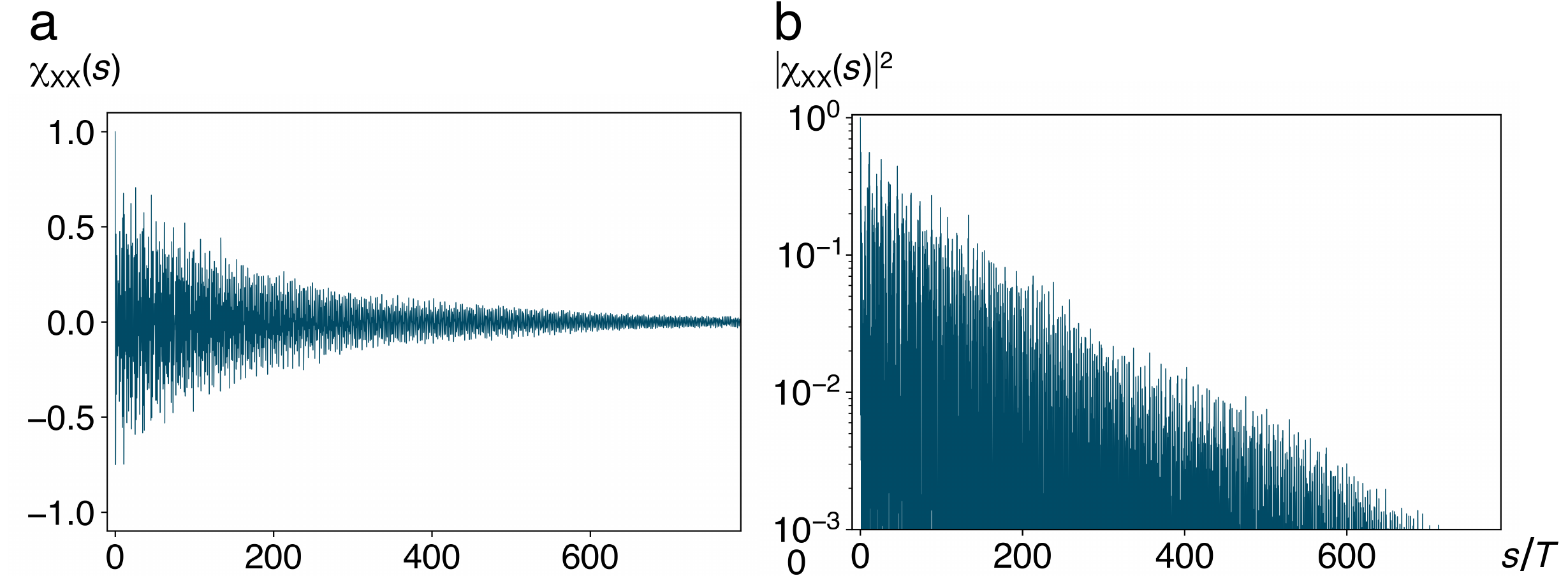}
\caption{
Autocorrelation $\chi_{XX}(s)$, cf.\ Eq.\ (\protect\ref{xxautocorr}), of the position $X(t)$
of the central system (a). Despite averaging over ensembles of initial conditions of the bath
as well as the absolute time in times series of $X(t)$, oscillations with the frequency $\Omega$
of harmonic motion around the stable equilibria survive longer than bath oscillations proper.
A semilogarithmic plot of the square $\left| \chi_{XX}(s) \right|^2$ (b) confirms their exponential decay.
Time axes in units of $T = 2\pi / \Omega$ as in Fig.\ \protect\ref{figjumps}. Other parameter values are
$a = 2$, $b = 1$, $g = 0.5$, $E_{\rm{bath}} = 0.1$, $M = 1$, $m = 0.1$, $N = 15$,
$\omega_{\rm{co}} = 4$.
}
\label{figautocorr}
\end{center}
\end{figure}   

An example of the time-averaged autocorrelation $\chi_{XX}(s)$ is plotted in
Fig.\ \protect\ref{figautocorr}a. Superposed on the long-term exponential decay of the envelope,
we observe rapid oscillations of the autocorrelation, with the frequency $\Omega$ of harmonic
motion around the stable equilibria on the bottom of each well. They decay much slower than
the fluctuations of the heat bath. In panel (b), we show a semilogarithmic of the square
$\left| \chi_{XX}(s) \right|^2$, circumventing negative values of $\chi_{XX}(s)$, as direct evidence
of the exponential decay of the autocorrelation.
 
The process of relaxation of the central system towards a stationary state in one of
the wells should be reflected in a characteristic time dependence of the entropy
of the subsystems. While the total entropy of the system double well plus
environment is conserved under canonical transformations, the sum of partial entropies of
subsystems may vary. In terms of the reduced probability density of the double well,

\begin{equation} \label{sredudens}
\rho_{\rm{S}}(\mathbf{R},t) = \int {\rm{d}}^{2N} r \, \rho(\mathbf{R},\mathbf{r}) =
\int {\rm{d}} p_N \int {\rm{d}} x_N \cdots \int {\rm{d}} \, p_1 \int {\rm{d}} x_1 \, \rho(\mathbf{R},\mathbf{r},t),
\end{equation}

\noindent
the partial entropy in this subsystem is given by \cite{Bri56,KS04}

\begin{equation} \label{sentropy}
S_{\rm{S}} = -c \int {\rm{d}}^2 R \, \rho_{\rm{S}}(\mathbf{R}) \ln(\rho_{\rm{S}}(\mathbf{R}) \Delta A)
\end{equation}

\noindent
where $\Delta A$ is the symplectic area of a minimal phase-space cell resolved by the input data,
and we choose $c = 1/\ln(2)$, measuring entropy in units of bits.
We evaluate the entropy by launching a set of $K$ trajectories, initially concentrated within
a single bin $\Delta A = \Delta P \Delta X$ of a discretized phase space, and counting
the number of trajectories found at time $t$ in each bin at $(P_\lambda,X_\mu)$,
$\lambda = 1,\ldots,L$, $\mu = 1,\ldots,M$, to determine probabilities $p_{\lambda,\mu}(t)$.
The entropy is then calculated as

\begin{equation} \label{sentrodisc}
S_{\rm{S}}(t) = -c \sum_{\lambda = 1}^L \sum_{\mu = 1}^M
p_{\lambda,\mu}(t) \ln\bigl(p_{\lambda,\mu}(t)\bigr).
\end{equation}

\begin{figure}[h!]
\begin{center}
\includegraphics[width=11 cm]{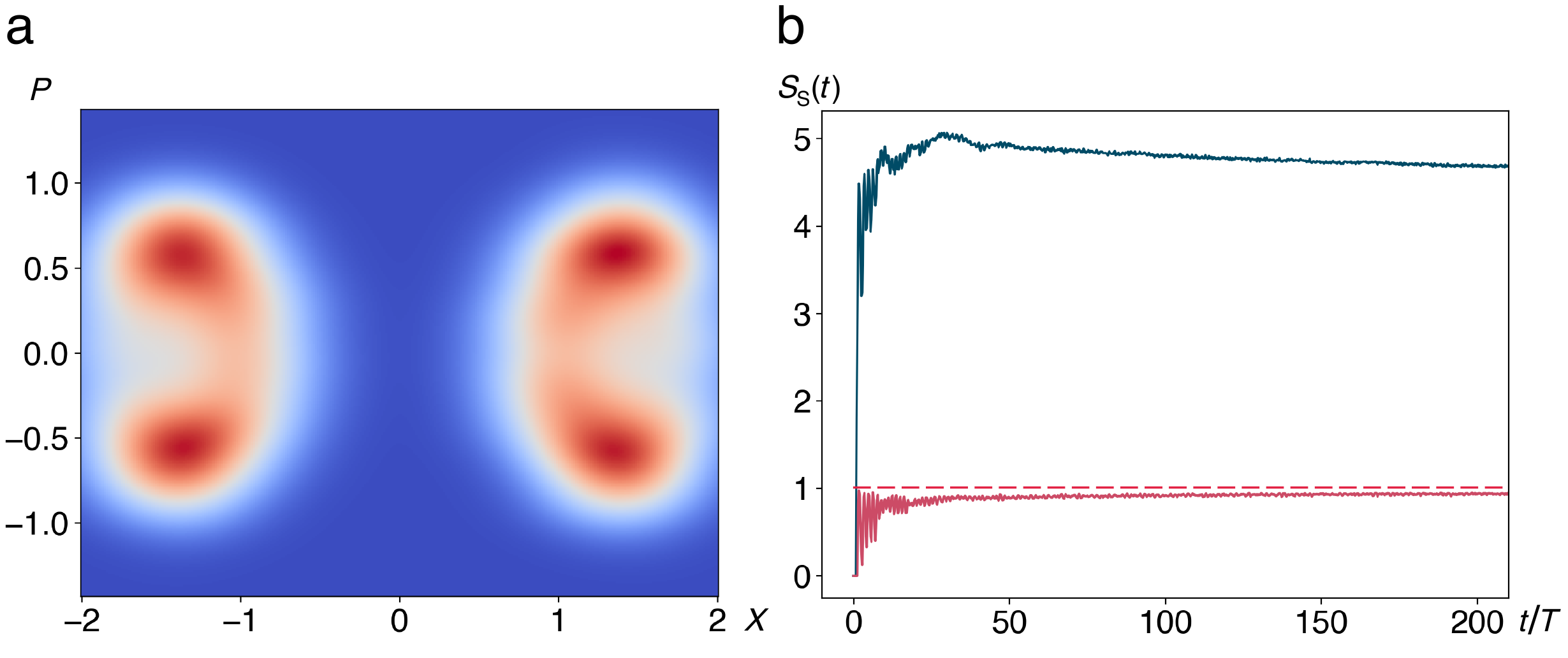}
\caption{
(a) Snapshot of the reduced density $\rho_{\rm{S}}(P,X,t)$, Eq.\ (\protect\ref{sredudens}),
at $t = 200 T$, $T = 2\pi/\Omega$, featuring the bimodal distribution with one peak in each of
the two minima of the double well. Colour code ranges from zero (blue) through moderate (white)
through high positive values (red). (b) Time evolution of the partial entropy $S_{\rm{S}}(t)$ of the central
system, Eq.\ (\protect\ref{sentrodisc}), initially prepared at rest on top of the barrier, calculated with
different degrees of resolution of phase space. For a large bin size $\Delta A = 3$ (red),
only the splitting of the asymptotic distribution into two peaks is resolved, resulting in
an asymptotic entropy $S_{\rm{S}}(t) \to 1 {\rm{bit}}$ (dashed horizontal line).
For $\Delta A = 0.5$ (blue), the entropy is dominated by the finite width of the two peaks,
approaching an asymptote $\lim_{t \to \infty} S_{\rm{S}}(t) \gg 1 {\rm{bit}}$.
Time axis in units of $T$ as in Fig.\ \protect\ref{figjumps}. Other parameter values are
$a = 2$, $b = 1$, $g = 0.1$, $E_{\rm{bath}} = 0.1$, $M = 1$, $m = 0.1$, $N = 15$,
$\omega_{\rm{co}} = 4$.
}
\label{figentropy}
\end{center}
\end{figure}

\noindent
By construction, $S_{\rm{S}}(0) = 0$. Time series of the partial entropy in the degree of freedom
of the double well are presented in Fig.\ \protect\ref{figentropy}. If the phase-space resolution
$\Delta A$ is chosen comparable to the area of the two maxima of the bimodal asymptotic
density distribution (Fig.\ \protect\ref{figentropy}a), the entropy approaches an asymptote of
1 bit (red curve and dashed horizontal line in Fig.\ \protect\ref{figentropy}b). For a higher
resolution (blue curve in Fig.\ \protect\ref{figentropy}b), the major part of the entropy is
contributed by the nonzero widths of the two peaks, leading to a value far above 1 bit.

Calculating the partial entropy of the environment, to compare it with that of the central system,
requires discretizing a $2N$-dimensional phase space into reasonably fine bins,
a task that is unfeasible with the computer equipment available to us.

\begin{figure}[h!]
\begin{center}
\includegraphics[width=11 cm]{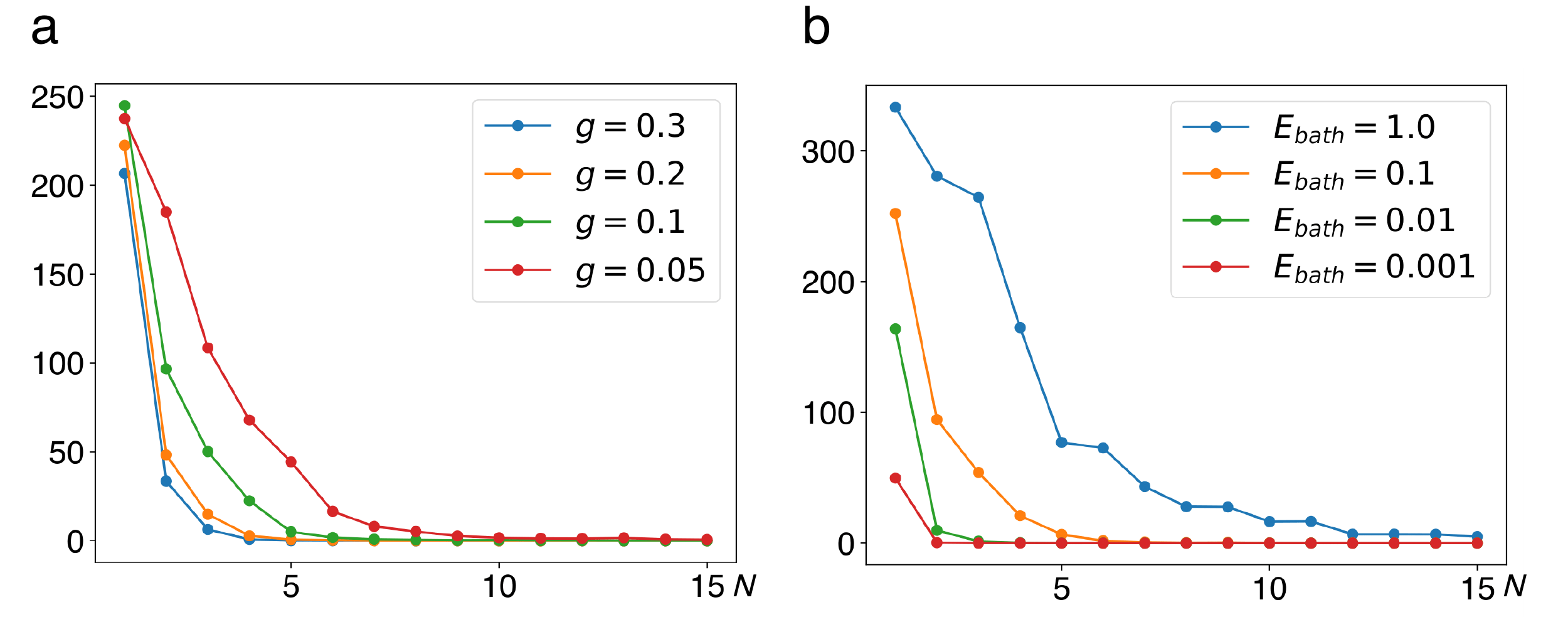}
\caption{
Frequency of jumps of the central system as a function of the number $N$ of bath modes,
varying the global coupling $g$ (a) and the total energy in the bath $E_{\rm{bath}}$ (b).
The number of jumps occurring within 100 periods $T = 2\pi / \Omega$ of unperturbed
oscillations around the stable equilibria of the double well has been averaged over
ensembles of 100 initial conditions of the heat bath oscillators. Other parameter values are
$a = 2$, $b = 1$, $E_{\rm{bath}} = 0.1$, $M = 1$, $m = 0.1$, $\omega_{\rm{co}} = 4$.
}
\label{figjumpfreq}
\end{center}
\end{figure}   

Finally, a simple direct criterion for the relaxation of a bistable system to a stable equilibrium
is the frequency of jumps between the two wells. They require a kinetic energy of the order
of the barrier height to be concentrated in the central degree of freedom, that is, an
exceptionally strong fluctuation. Therefore they become less and less likely as
the number of oscillators in the environment increases. We present evidence for
this tendency in Fig.\ \protect\ref{figjumpfreq}, plotting the number of jumps, accumulated over
a constant measurement period $\Delta t = 100 T$, with $T = 2\pi / \Omega$,
as a function of the number $N$ of modes in the heat bath,
varying the global coupling $g$, cf.\ Eq.\ (\ref{couplings}) (Fig.\ \protect\ref{figjumpfreq}a) and 
the total energy in the bath $E_{\rm{bath}}$, cf.\ Eq.\ (\ref{etotal}) (Fig.\ \protect\ref{figjumpfreq}b).
As is to be expected, the sojourn time in either minimum grows with increasing coupling
strength and with decreasing energy in the bath.

\subsection{Amplified fluctuations: randomness in the approach to an asymptotic state}\label{sec43}

Even if the time from the initial relaxation into one of the two minima till the next
jump to the other side and between subsequent jumps diverges with increasing size
of the bath, the Poincar\'e recurrence theorem \cite{MM60} implies that the system
will return infinitely often to a state within an $\epsilon$-neighbourhood of its initial state
(on top of the barrier) for every $\epsilon \in \mathbb{R}^+$. In the present context,
such a near recurrence occurs every time the system passes over the barrier, moving
from one well to the other. In this sense, there is no such thing as a ``final state'' of
a double well coupled to a finite heat bath. Notwithstanding, the time from
one jump to the next, thus between two subsequent recurrences, rapidly exceeds
every physically relevant timescale. Moreover, in the same guise of a realistic
modelling of a system embedded in its environment, with increasing time,
weaker couplings to remoter systems, hence larger environments, have to be
taken into account, leading to a further stabilization of the state the system
had relaxed to for the first time. These arguments justify talking of a ``final state''
in a heuristic sense.

\begin{figure}[h!]
\begin{center}
\includegraphics[width=11 cm]{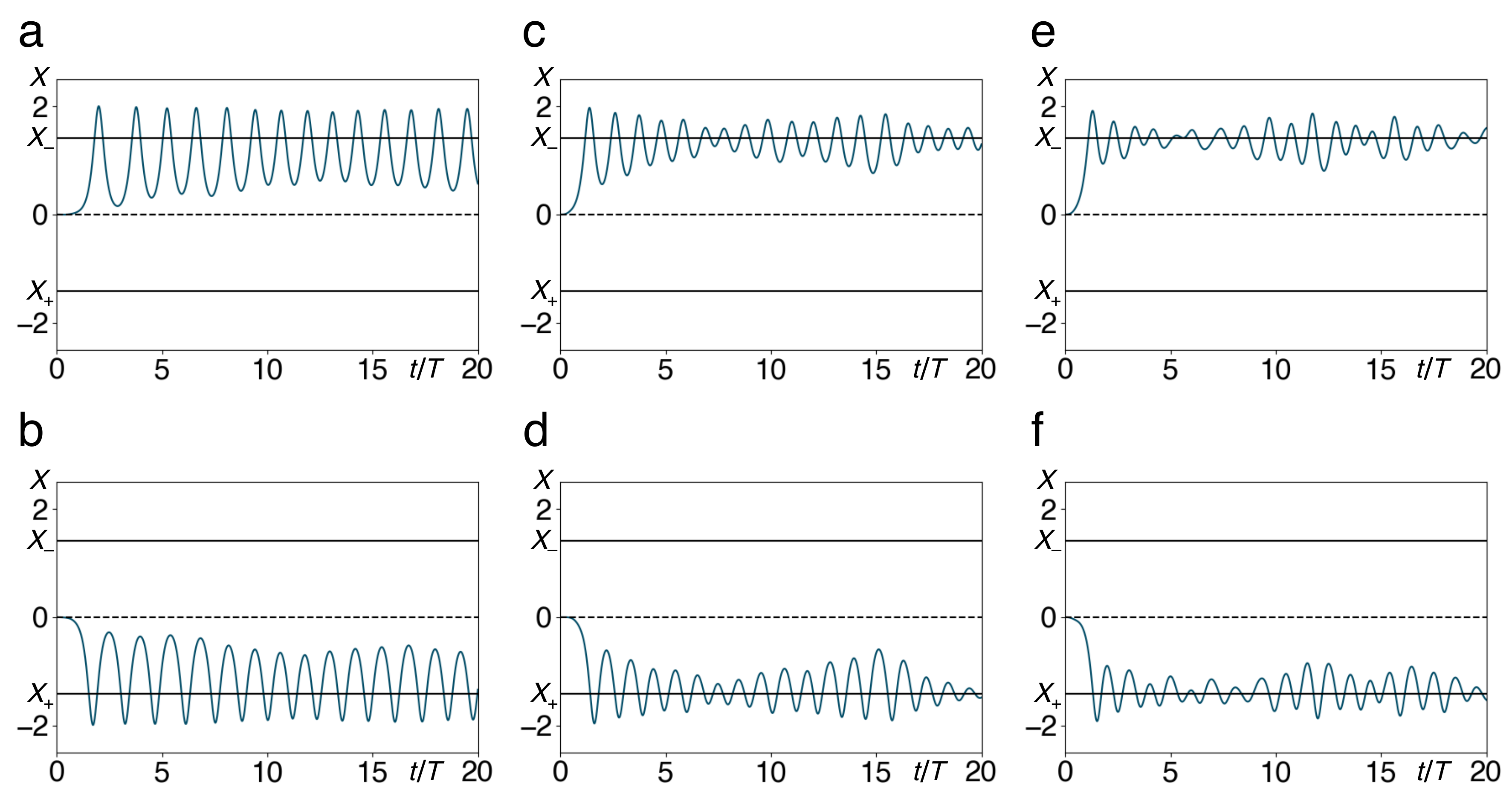}
\caption{
Sample trajectories of the quartic double well coupled to $N$ harmonic oscillators,
Eqs.\ (\protect\ref{sseehamiltonian}) to (\protect\ref{sepotential}),
falling from an initial state at rest on top of the barrier, $\bigl(P(0),X(0)\bigr) = (0,0)$,
into the left or the right well, plotted as traces position vs.\ time.
Time axes in units of $T = 2\pi / \Omega$ as in Fig.\ \protect\ref{figjumps}.
The stable equilibria $X = X_\pm$ and the unstable equilibrium at $X = 0$ of
the double well are marked by full and dashed horizontal lines, resp.
Parameter values are $g = 0.1$ (a,b), $0.3$ (c,d), $0.5$ (e,f), and $a = 2$, $b = 1$,
$E_{\rm{bath}} = 0.1$, $M = 1$, $m = 0.1$, $\omega_{\rm{co}} = 4$,
$N = 15$. Frequencies and initial conditions of the bath oscillators
have been drawn from random distributions, see text.
}
\label{figrelaxation}
\end{center}
\end{figure}   

In Fig.\ \protect\ref{figrelaxation}., we present a few example trajectories which, starting from
a state at erst at the top of the barrier, eventually fall into one of the two wells and then merely
fluctuate around the corresponding minimum. As the statistics of jumps in the foregoing
subsection already indicates, the amplitude of these fluctuations diminishes with
increasing size of the heat bath and with the coupling to it, but increases
with the total energy in the bath, that is, in a thermodynamical context, with
increasing temperature. These trajectories are reproducible: Prepared in
the same initial states of the double well and in particular of the bath oscillators,
the system approaches the same quasi-stationary state in the same way.

\begin{figure}[h!]
\begin{center}
\includegraphics[width=11 cm]{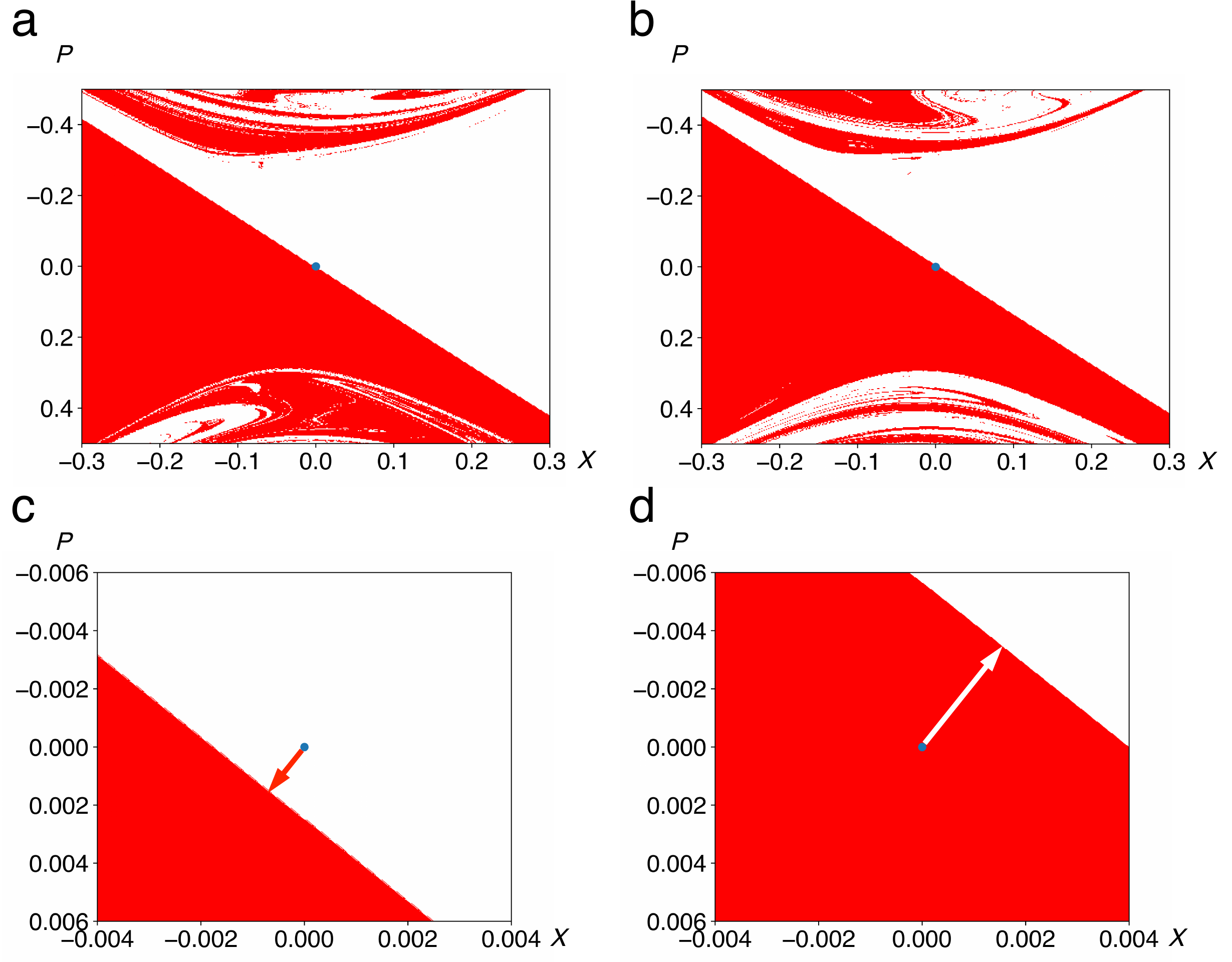}
\caption{
Basins of attraction of the two stable equilibria of the double well as in
Fig.\ \protect\ref{figdoublewell}c, but for different specific initial conditions of the bath.
Colours (red vs.\ white) indicate the side the system approaches in its initial relaxation.
Panels (a) and (b) are total views of the central part of the basins, showing their
disturbance by the initial condition of the bath. Close-ups (c,d) demonstrate the shift
of the boundary (arrows), away from the origin $\bigl(P(0),X(0)\bigr) = (0,0)$ (blue dot),
if the bias imparted by the bath lets the system fall from the top of the barrier into
the right (c) or the left well (d). Parameter values are $a = 2$, $b = 1$, $g = 0.1$,
$E_{\rm{bath}} = 0.1$, $M = 1$, $m = 0.1$, $N = 15$, $\omega_{\rm{co}} = 4$.
}
\label{figbasins}
\end{center}
\end{figure}   

Figure \protect\ref{figbasins} provides evidence for this scenario from a different point of view.
We depict the basins of attraction of the two wells (left well red, right well white) in the phase space
of the central system, now defined by the side the system approaches in its first relaxation,
\emph{keeping the initial condition of the bath fixed}. That means that the boundary no
longer passes exactly through the top of the barrier, as it does for the Newtonian equation of motion
with dissipation, Eq.\ (\ref{dwnewton}), see Fig.\ \protect\ref{figdoublewell}c. If for a given initial
state of the bath, the central system falls from the top of the barrier into, say, the left well,
this implies that an opposite bias has to be imposed on the initial condition of the
central system, for example an initial position slightly to the right of the barrier top
or a small positive momentum, to compensate for the initial bias of the bath.
In this way, the boundary is shifted towards positive position or momentum, into the
first quadrant of the phase space of central system (white arrow in Fig.\ \protect\ref{figbasins}d),
and vice versa if from $\bigl(P(0),X(0)\bigr) = (0,0)$, it falls into the right well
(red arrow in Fig.\ \protect\ref{figbasins}d).

Panels (a) and (b) of Fig.\ \protect\ref{figbasins} show the shape of the basins within a relatively large
phase-space domain. The self-similar patterns superposed on the smooth Yin-and-Yang shape of
Fig.\ \protect\ref{figdoublewell}c reflect the nonlinear nature of the dynamics as well as the random character of the initial state of the bath. Panels (c) and (d) show close-up views of the basin
boundary, close to the origin $(P,X) = (0,0)$, for two different initial conditions of the bath.
The shift of away from the origin is obvious. 

\section{Conclusions}\label{sec5}

With the project presented in this report, we have explored new ground
in several respects. By contrast to the theory of deterministic chaos, we here study
the origin of randomness in \emph{discrete} time series, such as those generated
by games of luck, in a deterministic dynamics. We substantiate our approach by constructing
a detailed model of a bistable system interacting with a many-body environment,
a quartic double well coupled to a bath comprising only a finite number of
harmonic oscillators, which evolves in time as a closed Hamiltonian system, thus
conserving information and energy.

Numerical solutions of the equations of motion reveal a rich dynamical scenario:
For a single harmonic oscillator coupled to the double well, we observe Hamiltonian
chaos emerging from integrable behaviour as predicted by the KAM theorem.
Increasing the number of bath modes, the system comes closer and closer
to an irreversible time evolution, replacing chaotic dynamics by relaxation into states
that remain stable on increasingly long time scales. Being bistable and symmetric
under spatial reflection, the long-time dynamics comprises two attractors,
left and right minimum, which are approached with equal probability 0.5. Which one
is reached, starting from an unbiased initial state of the central system on top of the barrier
separating the minima, is reproducibly determined by the initial state of
the environment.

With this behaviour, our model amplifies microscopic fluctuations to macroscopically
observable randomness. Unlike Brownian motion, however, this stochastic process
does not become manifest as a continuous quivering but as a stable discrete variable,
a random binary number or a sequence of them if the trial is repeated. It keeps a lasting memory,
encoding the initial state of the environment in a single bit. In this way, it reconciles the random
outcomes of this toppling pencil experiment (analogous to tossing a coin) with two
fundamental symmetries: It identifies the environment as the source of the entropy
generated by the binary random sequence that violates the conservation of
information in the macroscopic degree of freedom alone, and it explains how
the parity symmetry of potential and initial state of the bistable system is broken
by a microscopic bias in the initial state of the environment.

We hope that our work may serve as a template for further studies of discrete
stochastic phenomena in systems that allow for a classical or semiclassical
description, down to molecular physics. It remains an open question allows
for any kind of conclusion concerning randomness in quantum systems.
As a heuristic quantization of a double well coupled to a finite heat bath,
we presently investigate the spin-boson model with a finite number of boson modes
to provide some insight in this respect.

\section*{Funding}

A B.Sc.\ studentship for SPM, provided by Universidad Nacional de Colombia
(project code 37108), is gratefully acknowledged.

\section*{Acknowledgments}
We enjoyed inspiring discussions with Frank Gro\ss mann,
Oscar Rodr\'\i guez, Walter Strunz, and Carlos Viviescas. One of us (TD) thanks for
the hospitality extended to him by the Institute for Theoretical Physics at Technical University of Dresden
(Dresden, Germany) during various research stays, where part of the work reported here has been performed,
and is grateful for online access to their journal library granted to him by the Max Planck Institute for the Physics
of Complex Systems (MPIPKS, Dresden, Germany).

\section*{Appendix: Discretizing frequencies according to a given spectral density} \label{specdisdens}

We would like to construct a sequence of discrete frequencies $\omega_n$,
$n = 1,\ldots,N$, that comply with a given density of states,

\begin{equation} \label{freqdens}
\rho_\omega(\omega) \, {\rm{d}} \omega :=
\sum_{\scriptstyle n \atop \scriptstyle \omega \le \omega_n \le \omega + {\rm{d}} \omega}
\delta(\omega - \omega_n).
\end{equation}

\noindent
In this task, we have to cope with the difficulty that this functional dependence is defined with respect
to the frequency $\omega$, not to the actual independent variable, which in this context is
a discrete index $n$, such as a quantum number counting energy eigenstates of a fictitious
Hamiltonian. This requires to determine the functional dependence $\omega(x)$ on
an independent variable $x$, continuous to begin with, so that, with an equidistant discretization
of this variable,

\begin{equation} \label{xdisc}
x_n = x_0 + n \Delta x, \quad \Delta x = \frac{x_N - x_0}{N}, \quad n = 1,\ldots, N,
\end{equation}

\noindent
the frequencies $\omega_n = \omega(x_n)$ satisfy the required density of states
$\rho_\omega(\omega) = \rho_\omega\bigl(\omega(x_n)\bigr)$.

The function relating these two quantities, the frequency as a function of $x$

\begin{equation} \label{xtoomega}
\mathbb{R}^+ \ni x \mapsto \omega(x) \in \mathbb{R}^+,
\end{equation}

\noindent
results in a level density depending on $x$,

\begin{equation} \label{xdens}
\rho_\omega(x) \, {\rm{d}} \omega := {\rm{d}} n_\omega(x),
\end{equation}

\noindent
where $n_\omega(x)$ counts the number of discrete frequencies $\omega(x_n)$ found
between some lower bound $x_0$ and $x$. With the discretization (\ref{xdisc}),
assuming a sufficiently smooth function $\omega(x)$, it can be expressed
in terms of the frequency step size or nearest-neighbour level separation,

\begin{equation} \label{stepsize}
\sigma_n := \omega(x_n) - \omega(x_{n-1})
\approx \left.(x_n - x_{n-1}) \frac{{\rm{d}} \omega(x)}{{\rm{d}} x} \right\vert_{x=x_n}
= \Delta x \omega'(x_n),
\end{equation}

\noindent
as

\begin{equation} \label{invstepsize}
\rho_\omega(x_n) \approx \frac{1}{\omega_n - \omega_{n-1}}
= \frac{1}{\sigma_n} \approx \frac{1}{\Delta x \omega'(x_n)}.
\end{equation}

\noindent
Taking the inverse of Eq.\ (\ref{invstepsize}) and setting $\Delta x = 1$,
we obtain a general functional equation the function $\omega(x)$ has to satisfy,

\begin{equation} \label{functional}
\omega'(x) 
= \frac{1}{\rho_\omega\bigl(\omega(x)\bigr)}.
\end{equation}

\noindent
Equation (\ref{functional}) can also be understood as a direct consequence of the relation
$\rho_\omega(\omega) {\rm{d}} \omega = \rho_x(x) {\rm{d}} x$, with $\rho_x(x) = {\rm{const}} = 1$,
implied by Eq.\ (\ref{xdisc}) with $\Delta x = 1$.

As an example, consider the algebraic density

\begin{equation} \label{subsuperohmic}
\rho_\omega(\omega) = \Delta \omega^{-s-1} \omega^s, \quad \omega \ll \omega_{\rm{co}},
\end{equation}

\noindent
(without cutoff, therefore not normalized), which includes the subohmic ($s < 1$),
the ohmic ($s = 1$) and superohmic ($s > 1$) regimes of solid-state physics \cite{LC+87,HW+19}.
$\Delta \omega$ is a frequency scale introduced to make sure that $\rho_\omega(\omega)$ has
the correct dimensionality $\omega^{-1}$. The differential equation,

\begin{equation} \label{differeq}
\omega'(x) = \Delta \omega^{s+1} \bigl(\omega(x)\bigr)^{-s}
\end{equation}

\noindent
(substituting Eq.\ (\ref{subsuperohmic}) into Eq.\ (\ref{functional})) is solved by

\begin{equation} \label{subsupspec}
\omega_n = (s + 1)^{\frac{1}{s+1}} \Delta \omega \, n^{\frac{1}{s+1}} ,
\end{equation}

\noindent
so that the inverse spectral step size as a function of $n$ becomes

\begin{equation} \label{subsupinvstep}
\omega'_n = (s + 1)^{\frac{-s}{s+1}}  \Delta \omega \, n^{\frac{-s}{s+1}}.
\end{equation}

\noindent
Resolving Eq. (\ref{subsupspec}) for $n(\omega)$, the level number as a function of the frequency,

\begin{equation} \label{levelnumber}
n(\omega) = \left\lceil \frac{1}{s+1} \left(\frac{\omega}{\Delta \omega} \right)^{s+1} \right\rceil,
\end{equation}

\noindent
substituted in Eq.\ (\ref{subsupinvstep}), allows us to check Eq.\ (\ref{functional}) for the spectral density (\ref{subsuperohmic}),

\begin{equation} \label{checksubsuperohmic}
\rho_\omega(\omega) = \frac{1}{\omega'\bigl(n(\omega)\bigr)}
= (s + 1)^{\frac{s}{s+1}} \frac{\bigl(n(\omega)\bigr)^{\frac{s}{s+1}}}{\Delta \omega}
= (\Delta \omega)^{-s-1} \omega^s.
\end{equation}

\begin{figure}[h!]
\begin{center}
\includegraphics[width=13 cm]{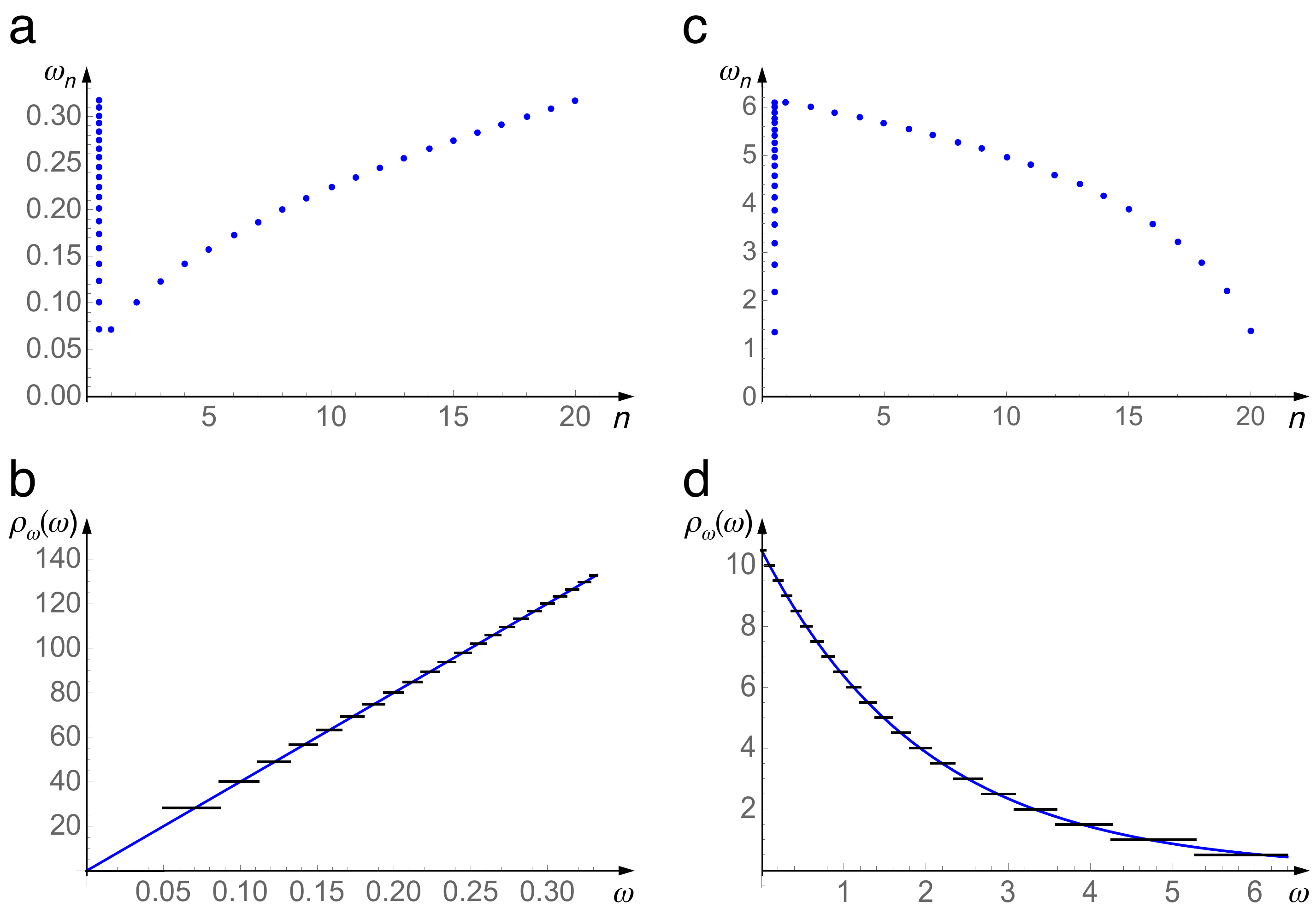}
\caption{
Results of the spectral discretization procedure detailed in Appendix \protect\ref{specdisdens},
for an Ohmic spectral density without cutoff, Eq.\ (\protect\ref{ohmicspec}) (panels (a), (b)),
and a pure exponential cutoff, Eq.\ (\protect\ref{expcutoff}) (c,d). Top (a,c): discrete frequencies
$\omega_n$ vs.\ level index $n$, projected onto a line near the vertical axis to visualize
the density. Bottom (b,d): Target spectral density (full blue curves) compared to the inverse
step size (black stairs), Eq.\ (\protect\ref{invstepsize}), generated by the procedure.
}
\label{figspecdisdens}
\end{center}
\end{figure}

\noindent
In particular, for $s = 1$ (Ohmic spectral density), see Fig.\ \ref{figspecdisdens}a,b,

\begin{equation} \label{ohmicspec}
\omega_n = \Delta \omega \sqrt{2n}.
\end{equation}

\noindent
An exponential cutoff for large frequencies,

\begin{equation} \label{expcutoff}
\rho_\omega(\omega) = \frac{1}{\omega_{\rm{co}}} \exp\left(\frac{\omega}{\omega_{\rm{co}}}\right),
\end{equation}

\noindent
is obtained with the discrete spectrum, see Fig.\ \ref{figspecdisdens}c,d,

\begin{equation} \label{discutoff}
\omega_n = - \omega_{\rm{co}} \ln(N_{\rm{co}} - n), \quad \omega \gg \omega_{\rm{co}},
\end{equation}

\noindent
Equations (\ref{subsupspec}) and (\ref{discutoff}) have to be matched adapting the free parameters
$\omega_{\rm{co}}$ and $N_{\rm{co}}$, with $N_{\rm{co}} > N$, accordingly. A solution
interpolating between Eqs.\ (\ref{ohmicspec}) and (\ref{discutoff}) is formally possible but
not in closed analytical form.

\end{document}